\documentclass[usenatbib]{aastex631}
\usepackage{array}
\usepackage{amsbsy}
\usepackage{graphicx}

\makeatletter

\usepackage[nointegrals]{wasysym}

\makeatother
\begin{document}

\title{Effects of Emerging Bipolar Magnetic Regions in Mean-field Dynamo
Model of Solar Cycles 23 and 24}

\shorttitle{Dynamo Model of Solar Cycles 23 and 24} \shortauthors{Pipin,
Kosovichev, \& Tomin}

\correspondingauthor{A. G. Kosovichev} 

\email{alexander.g.kosovichev@njit.edu}

\author{V.V.~Pipin}

\affil{Institute of Solar-Terrestrial Physics, Russian Academy of Sciences,
Irkutsk, 664033, Russia}

\author{A.G. Kosovichev}

\affiliation{New Jersey Institute of Technology, NJ 07102, USA} 

\author{V.E.~Tomin}

\affil{Institute of Solar-Terrestrial Physics, Russian Academy of Sciences,
Irkutsk, 664033, Russia}
\begin{abstract}
We model the physical parameters of Solar Cycles 23 and 24 using a
nonlinear dynamical mean-field dynamo model that includes the formation
and evolution of bipolar magnetic regions (BMR). The Parker-type dynamo
model consists of a complete MHD system in the mean-field formulation:
the 3D magnetic induction equation, and 2D momentum and energy equations
in the anelastic approximation. The initialization of BMR is modeled
in the framework of Parker's magnetic buoyancy instability. It defines
the depths of BMR injections, which are typically located at the edge
of the global dynamo waves. The distribution with longitude and latitude
and the size of the initial BMR perturbations are taken from the NOAA
database of active regions. The modeling results are compared with
various observed characteristics of the solar cycles. Only the BMR
perturbations located in the upper half of the convection zone lead
to magnetic active regions on the solar surface. While the BMR initialized
in the lower part of the convection zone do not emerge on the surface,
they still affect the global dynamo process. Our results show that
BMR can play a substantial role in the dynamo processes, and affect
the strength of the solar cycle. However, the data-driven model shows
that the BMR effect alone cannot explain the weak Cycle 24. This weak
cycle and the prolonged preceding minimum of magnetic activity were
probably caused by a decrease of the turbulent helicity in the bulk
of the convection zone during the decaying phase of Cycle 23. 
\end{abstract}


\section{Introduction}

The basic scenario for the hydromagnetic solar dynamo involves a cyclic
mutual transformation of the toroidal and poloidal magnetic fields
by means of differential rotation and cyclonic convective motions
characterized by kinetic helicity \citep{Parker1955}. The key idea
of Parker suggests that the poloidal magnetic field of the Sun is
generated from rising loops of the toroidal magnetic field in the
deep convection zone, which are twisted around the radial direction
by turbulent cyclonic motions (``the alpha-effect''). The resulting
poloidal fields of the loops coalesce into a large-scale poloidal
magnetic field of a new solar cycle. The subsequent stretching of
the poloidal field by the differential rotation produces the toroidal
field (``the Omega-effect''). The whole 22-year cyclic process of
the magnetic field generation and transformation represents dynamo
waves, forming the magnetic butterfly diagram and reversing the magnetic
polarity of the Sun's global magnetic field. Occasional strands of
toroidal flux tubes emerge on the surface due to the magnetic buoyancy
instability \citep{Parker1955a,Parker1979} in the form of east-west
oriented Bipolar Magnetic Regions (BMR).

On the one hand, this scenario stems from the turbulent dynamo theory
and mean-field dynamo models \citep{Moffatt1978,Parker1979,Raedler1980},
which are the basis of our current understanding of the nature of
magnetic activity in astrophysical objects \citep{Brandenburg2005b,sh_su21}.
Global convective MHD simulations prove the validity of the basic
principles of the dynamo theory \citep{Guerrero2016a,Schrinner2011,Schrinner2011a,Warnecke2018,Warnecke2021}.
Naturally, the mean-field solar dynamo models consider the dynamo
process to be distributed over the convection zone and `shaped' into
the butterfly diagram of emerged BMR in the subsurface rotational
shear layer \citep{Brandenburg2005,Pipin2011a}.

On the other hand, the phenomenological scenario of Babcock-Leighton
(hereafter, BL) \citep{Babcock1961,Leighton1969}, based on observations
of magnetic field evolution on the solar surface, grew into a popular
flux-transport model of the solar cycles (see reviews by \citealp{Charbonneau2011,Brun2014,Dikpati2016a}).
In this scenario, the poloidal field is generated due to the latitudinal
tilt of BMR emerging on the solar surface. This field is transported
by the surface meridional circulation and turbulent diffusion to the
polar regions where it sinks in the interior and is amplified by the
differential rotation, creating a new toroidal field, which produces
emerging BMR. The meridional circulation speed controls the solar-cycle
duration. For these dynamo models, the surface magnetic activity,
which is often considered in the form of an empirical ``source term''
in the induction equation, is crucial.

In terms of the mean-field theory, the empirical Babcock-Leighton
source term calculated from the magnetic flux of BMR observed on the
solar surface can be considered as a near-surface alpha-effect \citep{Stix1974}.
From this point of view, the primary differences between Parker's
and Babcock-Leighton's scenarios are in the distribution of the alpha-effect
in the convection zone and the role of the turbulent magnetic diffusion
and meridional circulation in the magnetic flux transport. While Parker's
scenario assumes that the alpha-effect is distributed in the convection
zone and turbulent diffusion plays a major role in the formation of
migrating dynamo waves and considers the formation of BMR as a secondary
effect, in the Babcock-Leighton scenario, the BMR play a key role
providing the surface alpha-effect and magnetic flux transported by
the meridional circulation to the polar regions.

Theoretical models based on these scenarios have been successful in
explaining some observed properties of the solar cycles and magnetic
field evolution. The Babcock-Leighton flux-transport models explain
the magnetic flux emergence and transport observed on the solar surface
as well as the evolution of the polar magnetic field \citep{Dikpati2016}.
The recently developed self-consistent Parker-type mean-field model
\citep{Pipin2018b} explains the global magnetic structure and evolution
\citep{Pipin2021b}, as well as the dynamical processes, such as the
migrating zonal flows -- torsional oscillations \citep{Kosovichev2019}
and variations of the meridional circulation \citep{Getling2021},
observed by helioseismology. The model also explained the extended
solar-cycle phenomenon \citep{Pipin2019c}. This model determines
the 3D evolution of the large-scale vector magnetic field, coupled
with the equations describing large-scale flows and heat transport
in the axisymmetric 2D approximation. Thus, unlike in the Babcock-Leighton
models, the differential rotation and meridional circulation are not
input parameters - they are calculated together with the magnetic
field evolution. This allowed comparing the model results with the
helioseismic measurements.

Despite the recent advances, both types of dynamo models do not provide
a clear understanding of the physical mechanisms causing variations
of the solar amplitude and, thus, cannot provide robust solar-cycle
predictions. In the Babcock-Leighton models, the primary idea is that
the solar cycle strength is governed by spatial and temporal variations
of the BMR tilt, which affects the amount of magnetic flux transported
to the polar regions \citep[e.g.][]{Dikpati2016a}. On the other hand,
Parker-type models attempt to explain the solar-cycle amplitudes by
long- and short-term variations of the distributed alpha-effect and
associated non-linear processes in the deep convection zone \citep[e.g.][]{Pipin2020}.

To evaluate the role of BMR in these models, \citet{Pipin2022} developed
a 3D mean-field model, which includes the mechanism of magnetic flux
emergence due to the magnetic buoyancy instability resulting in the
formation of BMR on the surface. It was found that the generation
of the large-scale poloidal field via the emergence and evolution
of the solar bipolar magnetic regions can profoundly effect on the
global dynamo process. Thus, this new type of mean-field dynamo modeling
includes the basic features of both Parker's and Babcock-Leighton's
scenarios.

In this paper, we study the effects of BMR on the solar dynamo using
the mean-field dynamo model of \citet{Pipin2020} and the BMR formulation
of \citet{Pipin2022}. Our aim is to investigate the global parameters
of the mean-field dynamo for these solar cycles and evaluate the role
of BMR in the cycle properties. To model the surface magnetic activity,
we adopt the data of solar active regions from the NOAA Space Weather
Prediction Center for Solar Cycles 23 and 24. In particular, we study
the evolution of the polar magnetic field and investigate whether
the modeled BMR activity can explain the weak Cycle 24.

\section{Model formulation}

The model is formulated within the mean-field MHD (magnetohydrodynamics)
framework of \citet{Krause1980}. The basic details of the model are
given earlier by \citet{Pipin2019c} (hereafter PK19) and \citet{Pipin2022}
(hereafter, P22).

The magnetic field evolution is governed by the mean-field induction
equation: 
\begin{equation}
\partial_{t}\left\langle \mathbf{B}\right\rangle =\mathbf{\nabla}\times\left(\mathbf{\mathbf{\boldsymbol{\mathbf{\mathcal{E}}}}+}\left\langle \mathbf{U}\right\rangle \times\left\langle \mathbf{B}\right\rangle \right)\,,\label{eq:mfe}
\end{equation}
where $\mathbf{\mathcal{E}}=\left\langle \mathbf{u\times b}\right\rangle $
is the mean electromotive force; $\mathbf{u}$ and $\mathbf{b}$ are
the turbulent fluctuating velocity and magnetic field, respectively;
and $\left\langle \mathbf{U}\right\rangle $ and $\left\langle \mathbf{B}\right\rangle $
are the mean velocity and magnetic field. We assume that the averaging
is done over the ensemble of turbulent flows and magnetic fields.
We decompose the induction vector $\left\langle \mathbf{B}\right\rangle $
into the sum of the axisymmetric and non-axisymmetric parts, 
\begin{equation}
\left\langle \mathbf{B}\right\rangle =\overline{\mathbf{B}}+\tilde{\mathbf{B}},\label{eq:b0}
\end{equation}
where $\overline{\mathbf{B}}$ and $\tilde{\mathbf{B}}$ are the axisymmetric
and non-axisymmetric components of the large-scale magnetic field.
The mean electromotive force describes the turbulent effects on the
mean magnetic field evolution. It consists of two parts: 
\begin{eqnarray}
\mathcal{E}_{i} & = & \mathcal{E}_{i}^{(A)}+\mathcal{E}_{i}^{(\mathrm{BMR})},\label{eq:EMF-1}
\end{eqnarray}
where $\mathcal{E}_{i}^{(A)}$ is calculated analytically using the
double-scale approximation of \citet{Roberts1975} (see, e.g., \citealt{Kitchatinov1994,Pipin2008a}).
The phenomenological part $\mathcal{E}_{i}^{(\mathrm{BMR})}$describes
formations of the surface bipolar regions (BMR) from the large-scale
toroidal magnetic field.

The expression for $\mathcal{E}_{i}^{(A)}$ reads as follows, 
\begin{equation}
\mathcal{E}_{i}^{(A)}=\left(\alpha_{ij}+\gamma_{ij}\right)\left\langle B\right\rangle _{j}-\eta_{ijk}\nabla_{j}\left\langle B\right\rangle _{k},\label{eq:Ea}
\end{equation}
here, $\alpha_{ij}$ describes the turbulent generation of the magnetic
field by helical motions (the $\alpha$-effect), $\gamma_{ij}$ describes
the turbulent pumping, and $\eta_{ijk}$ is the eddy magnetic diffusivity
tensor. The $\alpha$-effect tensor includes the small-scale magnetic
helicity density contribution, i.e., the pseudo-scalar $\left\langle \chi\right\rangle =\left\langle \mathbf{a}\cdot\mathbf{b}\right\rangle $
(where $\mathbf{a}$ and $\mathbf{b}$ are the fluctuating vector-potential
and magnetic field, respectively), 
\begin{eqnarray}
\alpha_{ij} & = & C_{\alpha}\psi_{\alpha}(\beta)\alpha_{ij}^{(H)}+\alpha_{ij}^{(M)}\psi_{\alpha}(\beta)\frac{\left\langle \chi\right\rangle \tau_{c}}{4\pi\overline{\rho}\ell_{c}^{2}},\label{alp2d}
\end{eqnarray}
where the expressions of the kinetic helicity tensor $\alpha_{ij}^{(H)}$
and the magnetic helicity tensor $\alpha_{ij}^{(M)}$ are given by
\citet{Pipin2018b}. The radial profiles of the $\alpha_{ij}^{(H)}$
and $\alpha_{ij}^{(M)}$ depend on the mean density stratification,
profile of the convective RMS velocity $u_{c}$ and on the Coriolis
number $\Omega^{*}=2\Omega_{0}\tau_{c}$, where $\Omega_{0}$ is the
angular velocity of the star and $\tau_{c}$ is the convective turnover
time. The magnetic quenching function $\psi_{\alpha}(\beta)$ depends
on the parameter $\mathrm{\beta=\left|\left\langle \mathbf{B}\right\rangle \right|/\sqrt{4\pi\overline{\rho}u_{c}^{2}}}$.
Note that in the presence of the $\tilde{\mathbf{B}}$-field, the
$\alpha$ effect tensor becomes non-axisymmetric. It is caused by
the $\psi_{\alpha}(\beta)$-quenching and the magnetic helicity effects.

The magnetic helicity evolution follows the global conservation law
for the total magnetic helicity, $\left\langle \chi\right\rangle ^{(tot)}=\left\langle \chi\right\rangle +\left\langle \mathbf{A}\right\rangle \cdot\left\langle \mathbf{B}\right\rangle $,
(see, \citet{Hubbard2012,Pipin2013c,Brandenburg2018}): 
\begin{equation}
\left(\frac{\partial}{\partial t}+\boldsymbol{\left\langle \mathbf{U}\right\rangle \cdot\nabla}\right)\left\langle \chi\right\rangle ^{(tot)}=-\frac{\left\langle \chi\right\rangle }{R_{m}\tau_{c}}-2\eta\left\langle \mathbf{B}\right\rangle \cdot\left\langle \mathbf{J}\right\rangle -\mathbf{\nabla\cdot}\mathbf{\mathbf{\mathcal{F}}}^{\chi},\label{eq:helcon}
\end{equation}
where, we use ${\displaystyle 2\eta\mathbf{\left\langle b\cdot j\right\rangle }=\frac{\left\langle \chi\right\rangle }{R_{m}\tau_{c}}}$
\citep{Kleeorin1999}. Also, we introduce the diffusive flux of the
small-scale magnetic helicity density, $\mathbf{\mathbf{\mathcal{F}}}^{\chi}=-\eta_{\chi}\mathbf{\nabla}\left\langle \chi\right\rangle $,
and $R_{m}$ is the magnetic Reynolds number, we employ $R_{m}=10^{6}$.
Following the results of \citet{Mitra2010}, we put $\eta_{\chi}=\frac{1}{10}\eta_{T}$.
Further details about the turbulent dynamo effects can be found in
\citet{Pipin2022}(P22) and in the above-cited papers. In what follows,
we discard the advection of the total helicity by meridional circulation.
As a result, the amplitude of the polar magnetic field in the mean-field
model decreases in comparison to the standard case (cf, P22). One
purpose of this tuning is to get a stronger impact of the surface
BMR on the deep dynamo in the new model.

The turbulent pumping, which is expressed by the antisymmetric tensor
$\gamma_{ij}$, is important for reproducing the solar-like evolution
of the dynamo-generated magnetic field \citep{Warnecke2014,Warnecke2021}.
The formulation of $\gamma_{ij}$ for the solar-type mean-field dynamo
model was discussed by \citet{Pipin2018b}. We define it as follows,
\begin{eqnarray}
\gamma_{ij} & = & \gamma_{ij}^{(\Lambda\rho)}+\frac{\alpha_{\mathrm{MLT}}u_{c}}{\gamma}\mathcal{H}\left(\beta\right)\mathrm{\hat{r}_{n}\varepsilon_{inj}},\label{eq:pump0}\\
\gamma_{ij}^{(\Lambda\rho)} & = & 3\nu_{T}f_{1}^{(a)}\left\{ \left(\mathbf{\boldsymbol{\Omega}}\cdot\boldsymbol{\Lambda}^{(\rho)}\right)\frac{\Omega_{n}}{\Omega^{2}}\varepsilon_{\mathrm{inj}}-\frac{\Omega_{j}}{\Omega^{2}}\mathrm{\varepsilon_{inm}\Omega_{n}\Lambda_{m}^{(\rho)}}\right\} \label{eq:pump1}
\end{eqnarray}
where $\mathbf{\boldsymbol{\Lambda}}^{(\rho)}=\boldsymbol{\nabla}\log\overline{\rho}$,
$\mathrm{\alpha_{MLT}}=1.9$ is the mixing-length theory parameter,
$\gamma$ is the adiabatic exponent, $u_{c}$ is the RMS convective
velocity. In Eq.(\ref{eq:pump0}), the first term takes into account
the mean drift of the large-scale field due to the gradient of the
mean density, and the second term describes the magnetic buoyancy
effect. Function $\mathcal{H}\left(\beta\right)$ takes into account
the effect of magnetic tension. For small values of $\beta$, $\mathcal{H}\left(\beta\right)\sim\beta^{2}$,
and it saturates as $\beta^{-2}$ for $\beta\gg1$ (see P22 for details).

We assume that the large-scale flow is axisymmetric. It is decomposed
into a sum of the meridional circulation and differential rotation,
$\mathbf{\overline{U}}=\mathbf{\overline{U}}^{m}+r\sin\theta\Omega\left(r,\theta\right)\hat{\mathbf{\boldsymbol{\phi}}}$,
where $r$ is the radial coordinate, $\theta$ is the polar angle,
$\hat{\mathbf{\boldsymbol{\phi}}}$ is the unit vector in the azimuthal
direction, and $\Omega\left(r,\theta\right)$ is the angular velocity
profile. The angular momentum conservation and the equation for the
azimuthal component of large-scale vorticity, $\mathrm{\overline{\omega}=\left(\boldsymbol{\nabla}\times\overline{\mathbf{U}}^{m}\right)_{\phi}}$,
determine the differential rotation and meridional circulation: 
\begin{eqnarray}
\frac{\partial}{\partial t}\overline{\rho}r^{2}\sin^{2}\theta\Omega & = & -\boldsymbol{\nabla\cdot}\left(r\sin\theta\overline{\rho}\left(\hat{\mathbf{T}}_{\phi}+r\sin\theta\Omega\mathbf{\overline{U}^{m}}\right)\right)\label{eq:angm-1}\\
 & + & \boldsymbol{\nabla\cdot}\left(r\sin\theta\frac{\left\langle \mathbf{B}\right\rangle \left\langle B_{\phi}\right\rangle }{4\pi}\right),\nonumber 
\end{eqnarray}

\begin{eqnarray}
\mathrm{\frac{\partial\omega}{\partial t}} & \mathrm{=} & \mathrm{r\sin\theta\boldsymbol{\nabla}\cdot\left(\frac{\hat{\boldsymbol{\phi}}\times\boldsymbol{\nabla\cdot}\overline{\rho}\hat{\mathbf{T}}}{r\overline{\rho}\sin\theta}-\frac{\mathbf{\overline{U}}^{m}\overline{\omega}}{r\sin\theta}\right)}\label{eq:vort-2}\\
 & + & \mathrm{r}\sin\theta\frac{\partial\Omega^{2}}{\partial z}-\mathrm{\frac{g}{c_{p}r}\frac{\partial\overline{s}}{\partial\theta}}\nonumber \\
 & + & \frac{1}{4\pi\overline{\rho}}\left(\overline{\mathbf{B}}\boldsymbol{\cdot\nabla}\right)\left(\boldsymbol{\nabla}\times\overline{\mathbf{B}}\right)_{\phi}-\frac{1}{4\pi\overline{\rho}}\left(\left(\boldsymbol{\nabla}\times\overline{\mathbf{B}}\right)\boldsymbol{\cdot\nabla}\right)\overline{\mathbf{B}}{}_{\phi},\nonumber 
\end{eqnarray}
where $\hat{\mathbf{T}}$ is the turbulent stress tensor: 
\begin{equation}
\hat{T}_{ij}=\left(\left\langle u_{i}u_{j}\right\rangle -\frac{1}{4\pi\overline{\rho}}\left(\left\langle b_{i}b_{j}\right\rangle -\frac{1}{2}\delta_{ij}\left\langle \mathbf{b}^{2}\right\rangle \right)\right),\label{eq:rei}
\end{equation}
(see detailed description in \citealp{Pipin2019c}, PK19). Also, $\overline{\rho}$
is the mean density, $\mathrm{\overline{s}}$ is the mean entropy;
$\mathrm{\partial/\partial z=\cos\theta\partial/\partial r-\sin\theta/r\cdot\partial/\partial\theta}$
is the gradient along the axis of rotation. The mean heat transport
equation determines the mean entropy variations from the reference
state due to the generation and dissipation of large-scale magnetic
fields and flows \citep{Pipin2000}: 
\begin{equation}
\overline{\rho}\overline{T}\left(\frac{\partial\overline{\mathrm{s}}}{\partial t}+\left(\overline{\mathbf{U}}\cdot\boldsymbol{\nabla}\right)\overline{\mathrm{s}}\right)=-\boldsymbol{\nabla}\cdot\left(\mathbf{F}^{c}+\mathbf{F}^{r}\right)-\hat{T}_{ij}\frac{\partial\overline{U}_{i}}{\partial r_{j}}-\boldsymbol{\boldsymbol{\mathcal{E}}}\cdot\left(\nabla\times\left\langle \mathbf{B}\right\rangle \right),\label{eq:heat-1}
\end{equation}
where $\overline{T}$ is the mean temperature, $\mathbf{F}^{r}$ is
the radiative heat flux, $\mathbf{F}^{c}$ is the anisotropic convective
flux (see, PK19). The last two terms in Eq.~(\ref{eq:heat-1}) take
into account the convective energy gain and sink caused by the generation
and dissipation of large-scale magnetic and flow fields. The reference
profiles of mean thermodynamic parameters, such as entropy, density,
and temperature, are determined from the stellar interior model MESA
\citep{Paxton2011,Paxton2013}. The radial profile of the typical
convective turnover time, $\tau_{c}$, is determined by the MESA code,
as well. We assume that $\tau_{c}$ does not depend on the magnetic
field and global flows. The convective RMS velocity is determined
from the mixing-length approximation, 
\begin{equation}
\mathrm{u_{c}=\frac{\ell_{c}}{2}\sqrt{-\frac{g}{2c_{p}}\frac{\partial\overline{s}}{\partial r}},}\label{eq:uc-1}
\end{equation}
where $\ell_{c}=\alpha_{MLT}H_{p}$ is the mixing length, $\alpha_{MLT}=1.9$
is the mixing-length parameter, and $H_{p}$ is the pressure scale
height. Equation~(\ref{eq:uc-1}) determines the reference profiles
for the eddy heat conductivity, $\chi_{T}$, eddy viscosity, $\nu_{T}$,
and eddy diffusivity, $\eta_{T}$, as follows, 
\begin{eqnarray}
\chi_{T} & = & \frac{\ell^{2}}{6}\sqrt{-\frac{g}{2c_{p}}\frac{\partial\overline{s}}{\partial r}},\label{eq:ch-1}\\
\nu_{T} & = & \mathrm{Pr}_{T}\chi_{T},\label{eq:nu-1}\\
\eta_{T} & = & \mathrm{Pm_{T}\nu_{T}}.\label{eq:et-1}
\end{eqnarray}
\begin{figure}
\centering \includegraphics[width=0.7\columnwidth]{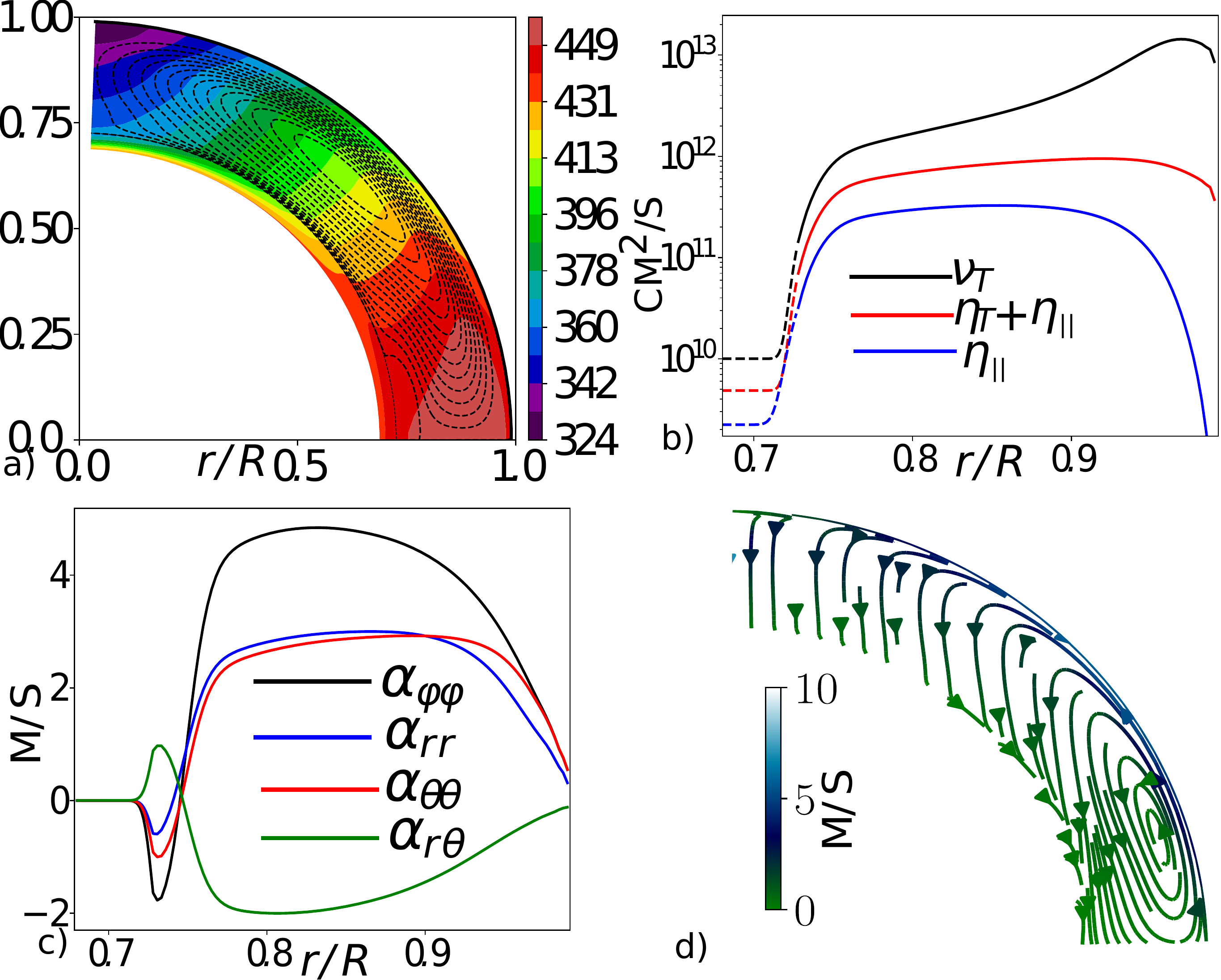} \caption{\label{fig1} a) The meridional circulation (streamlines) and the
angular velocity distributions; the magnitude of circulation velocity
is $\sim13$ m/s on the surface at 45$^{\circ}$ latitude; b) radial
dependencies of the total eddy magnetic diffusivity, $\eta_{T}+\eta_{||}$,
and its rotationally induced part, $\eta_{||}$, and the eddy viscosity
profile, $\nu_{T}$; c) the $\alpha$-effect tensor components as
a function of radius at 45$^{\circ}$ latitude; and d) the streamlines
of the toroidal magnetic field effective drift velocity caused by
the meridional circulation and the turbulent pumping effect. For post-processing
of modeling data and visualization in this and other figures, we use
\textsc{numpy/scipy} \citep{harris2020array,2020SciPyNMeth} together
with \textsc{matplotlib \citep{Hunter2007} } and \textsc{ pyvista
\citep{sullivan2019pyvista} }.}
\end{figure}

\begin{table}
\caption{\label{tab:bp}Basic parameters of the reference axisymmetric dynamo
model, see explanation in the text.}

\begin{tabular}{c|c}
\hline 
hydrodynamic \& heat transport & dynamo model \& boundary condition parameters\tabularnewline
\hline 
$\alpha_{\mathrm{MLT}}=1.9$,$\ell_{\mathrm{min}}=0.01R$, $\mathrm{Pr}_{T}=3/4$ & $C_{\alpha}=0.04$,$\mathrm{Pm}_{T}=10$, $\delta=0.999$, $B_{\mathrm{esq}}=5$G\tabularnewline
\hline 
\end{tabular}
\end{table}

{The solution of the heat transport equation determines the
mean entropy distribution. The entropy profile and the Coriolis number,
$\Omega^{*}=2\Omega_{0}\tau_{c}$ define the magnitude and distribution
of the turbulent effects in the convection zone. The profile of $\tau_{c}=\ell_{c}/u_{c}$
is taken from the output of the MESA code. For overshoot overshoot
region we assume that the intensity of the turbulent mixing decays
with exponent of $100$ from the bottom of the convection zone. We
assume that the bottom of overshoot region rotates as a solid-body
at the rate $\Omega_{0}=430$~nHz. The reference axisymmetric model
has only a few free parameters. We choose them to fit the reference
model into solar observations. }

{Firstly, for the distributed dynamo model the radial location
of the meridional circulation stagnation point is important. If the
stagnation point in the upper part of the convection zone then the
latitudinal turbulent pumping together with the meridional circulation,
and the Parker-Yoshimura law results into the solar-type dynamo waves
of the large-scale toroidal field (\citealt{Pipin2018b,Pipin21c}).
The location of the stagnation point and meridional circulation structure
depends on the $\Lambda$-effect profile in the solar convection zone.
It is determined by the mean-field expression obtained by \citet{Kitchatinov1993a}
and \citet{Kitchatinov2004}. In the standard mean-field models of
solar differential rotation, e.g., the model of \citet{Kitchatinov2005},
the $\Lambda$-effect is determined by the density gradient, turbulence
anisotropy and the Coriolis number profile of the solar convection
zone. The stagnation point of the meridional circulation in their
model is located near the bottom of the convection zone. Similar to
\citep{Pipin2018c} our model takee into account the radial gradient
of the convective turnover time, $\tau_{c}$ . This effects results
to inversion of the $\Lambda$-effect near the bottom of the convection
zone. It causes the origin of the second meridional circulation cell
(clockwise in the North) near the bottom of the convection zone. Also
the stagnation point of the main cell (anticlockwise in the North)
is shifted toward the top in this case According to the convection
zone properties given by the MESA code, $\tau_{c}$ increases sharply
towards the bottom of convection zone. To get the model with one circulation
cell and the stagnation point in the upper part of convection zone
we smooth the sharp variation of $\tau_{c}$ toward the bottom of
the convection zone using the following ansatz of \citet{Kitchatinov2017}:
}
\begin{equation}
\ell_{c}=\ell_{\mathrm{min}}+\frac{1}{2}\left(\ell_{c}^{(0)}-\ell_{\mathrm{min}}\right)\left[1+\mathrm{erf}\left(\frac{r-(r_{b}+\ell_{\mathrm{min}})}{R_{\odot}d}\right)\right],\label{lmin}
\end{equation}
where $\ell_{c}^{(0)}$ is the mixing-length parameter from the MESA
code, $r_{b}=0.728R_{\odot}$ is the radius of the bottom of the convection
zone, $d=0.02$. We use $\ell_{\mathrm{min}}$ as a control parameter
to model saturation of $\tau_{c}$ variations in the $\Lambda$-tensor.{
For }$\ell_{\mathrm{min}}=0${ we get the double cell meridional
circulation structure. Here, we use }$\ell_{\mathrm{min}}=0.01R${.
This model show one-cell meridional circulation structure with stagnation
point at $0.9R$ and the better agreement of the angular velocity
profile with helioseismology than the model with double cell meridional
circulation structure. Yet, the nonlinear model shows a weak second
cell near the bottom of the convection zone in this case (see, \citealt{Pipin2019c}).
The eddy heat conductivity and eddy viscosity coefficients. Also,
we employ a standard choice to relate eddy heat conductivity an eddy
viscosity $\mathrm{Pr}_{T}=3/4$. }

{Secondly, the full dynamo cycle period of $20$ years is reproduced
if $\mathrm{Pm}_{T}=11$. The critical threeshold of the dimensionless
$\alpha$-effect parameter is $C_{\alpha}=0.04$. Figure~\ref{fig1}
illustrates distributions of the angular velocity, meridional circulation,
the $\alpha$-effect, and the eddy diffusivity calculated for a nonmagnetic
model. The amplitude of the meridional circulation on the surface
is about $13$~m/s. We list the critical parameters in the Table
\ref{tab:bp}. The efffect of the boundary condition parameters will
be discussed in subsection 2.3.}

\subsection{Formation of Bipolar Magnetic Regions (BMR)}

Following ideas of \citet{Parker1955a,Parker1971}, the emergence
of the bipolar magnetic regions (BMR) is modeled by using the mean
electromotive force representing the magnetic buoyancy and twisting
effects acting on unstable parts of the axisymmetric magnetic field
as follows (P22): 
\begin{equation}
\mathcal{E}_{i}^{(\mathrm{BMR})}=\alpha_{\beta}\delta_{i\phi}\left\langle B\right\rangle _{\phi}+V_{\beta}\left(\hat{\boldsymbol{r}}\times\left\langle \mathbf{B}\right\rangle \right)_{i},\label{eq:ep}
\end{equation}
where the first term describes the $\alpha$-effect caused by the
BMR tilt, and the second term models the magnetic buoyancy instability.
The magnetic buoyancy velocity, $V_{\beta}$, includes the turbulent
and mean-field buoyancy effects \citep{Kitchatinov1992,Kitchatinov1993,Ruediger1995}:
\begin{eqnarray}
V_{\beta} & = & \frac{\alpha_{\mathrm{MLT}}u_{c}}{\gamma}\mathcal{H}\left(\beta_{m}\right)\xi_{\beta}(t,\boldsymbol{r})\label{eq:bu}
\end{eqnarray}
where function $\mathcal{H}\left(\beta\right)$ describes magnetic
tension, $\mathrm{\beta=\left|\left\langle \mathbf{B}\right\rangle \right|/\sqrt{4\pi\overline{\rho}u_{c}^{2}}}$,
and subscript 'm' marks unstable points. Function $\xi_{\beta}$,
defines the location and formation of the unstable part of the magnetic
field, 
\begin{eqnarray}
\!\xi_{\beta}^{(\pm)}\left(\boldsymbol{r},t\right)\! & = & C_{\beta}\negthinspace\tanh\mathrm{\left(\frac{t}{\tau_{0}}\right)}\exp\left(-m_{\beta}\left(\sin^{2}\!\left(\!\frac{\phi-\phi_{m}}{2}\!\right)\right.\right.\!\label{xib}\\
 &  & \left.\left.+\!\sin^{2}\!\left(\!\frac{\theta-\theta_{\mathrm{m}}}{2}\!\right)\!\right)\right)\psi(r,r_{m}^{(\pm)}),\,t<\delta\mathrm{t_{0}}\nonumber \\
 & = & 0,\,t>\delta t.\nonumber 
\end{eqnarray}
where $\psi$ is a kink type function of radius, 
\begin{eqnarray}
\psi(r,r_{m}^{(\pm)})\! & =\!\! & \frac{1}{4}\left(\!1\!+\!\mathrm{erf}\left(100\frac{\left(r-r_{m}^{(\pm)}\right)}{R}\right)\!\right)\label{eq:step}\\
 & \times & \left(\!1\!-\!\mathrm{erf}\left(100\frac{\left(r-(r_{m}^{(\pm)}+0.1R)\right)}{R}\right)\!\right),
\end{eqnarray}
where $r_{m}$ and $\theta_{m}$ are the radius and the latitude of
the toroidal magnetic field strength extrema in the convection zone.
The reader finds further details in P22 and the above-cited papers.
Then, the instability may act both near the bottom and near the top
of the convection zone. We handle these situations separately using
separate functions: $\xi_{\beta}^{(-)}\left(\boldsymbol{r},t\right)$
for the low half ($r_{m}^{(-)}<0.86R$) and $\xi_{\beta}^{(+)}\left(\boldsymbol{r},t\right)$
for the upper half ($r_{m}^{(+)}>0.86R$) of the convection zone.

Compared to P22, we modify the time evolution of the instability from
simple exponent to $C_{\beta}\tanh\mathrm{\left(\frac{t}{\tau_{0}}\right)}$
and calculate the unstable points in the whole convection zone. This
is similar to our earlier paper \citep{Pipin2015a}. Also, this formulation
allows for more flexible assimilation of the observational data of
solar active regions (see Sec.~\ref{subsec:Data-driven}). For consistency
with the results of P22, we use the parameter $C_{\beta}=180$. The
other parameters are the same as in P22, i.e., the emergence time,
$\delta t=5$ days, the BMR's growth rate, $\tau_{0}$=1 day; and
with $m_{\beta}=100$ we get the size of BMR about 10 heliographic
degrees. The perturbations are randomly initiated in time and longitude
in each hemisphere independently.

The radial and latitudinal positions of the unstable points are computed
using the instability parameter 
\begin{equation}
I_{\beta}=-r\frac{\partial}{\partial r}\log\frac{\left|\overline{B}\right|^{\zeta}}{\overline{\rho}},\label{eq:inst}
\end{equation}
where $\overline{B}$ is the strength of the axisymmetric toroidal
magnetic field and $\overline{\rho}$ is the density profile. The
power law index $\zeta=1$ corresponds to Parker's instability condition.
For a better match with the solar observations, we use $\zeta=1.2$.
Typically, $I_{\beta}>0$ at the upper edge of the dynamo wave (see
Fig.~3 in P22). This condition defines the depth of the unstable
perturbations. Therefore, the instabilities are initiated at the points
of the $\left|\overline{B}\right|I_{\beta}\left(r,\theta\right)$
maxima where $I_{\beta}\left(r_{m},\theta_{m}\right)>0$.

The $\alpha$-effect of the BMR is Eq.~\ref{eq:ep} is given as follows
\begin{equation}
\alpha_{\beta}=C_{\alpha\beta}\left(1+\xi_{\alpha}\right)\cos\theta V_{\beta}\psi_{\alpha}(\beta)\psi(r,r_{\alpha}).\label{eq:ab}
\end{equation}
Here, the amplitude of the $\alpha$-effect is determined by the local
magnetic buoyancy velocity. Parameter $\xi_{\alpha}$ controls random
fluctuations of the BMR's $\alpha$-effect. In the current formulation,
the $\alpha$-effect of the BMR is readily linked to the BMR tilt
\citep{Stix1974}. Parameter $C_{\alpha\beta}$ defines the mean tilt
(see, P22). The latitudinal dependence of this relationship is governed
by the factor $\cos\theta$, see Eq.~(\ref{eq:ab}). In addition,
we use a step-like function of Eq.~(\ref{eq:step}) to define the
radial extent of the BMR perturbations.

We model the randomness of the tilt using the parameter $\xi_{\alpha}$.
Similarly to the work of \citet{Rempel2005c} and \citet{Pipin2022},
the $\xi_{\alpha}$ evolution follows the Ornstein--Uhlenbeck process,
\begin{eqnarray}
\dot{\xi}_{\alpha} & = & -\frac{2}{\tau_{\xi}}\left(\xi_{\alpha}-\xi_{1}\right),\label{xia}\\
\dot{\xi}_{1} & = & -\frac{2}{\tau_{\xi}}\left(\xi_{1}-\xi_{2}\right),\nonumber \\
\dot{\xi}_{2} & = & -\frac{2}{\tau_{\xi}}\left(\xi_{2}-g\sqrt{\frac{2\tau_{\xi}}{\tau_{h}}}\right).\nonumber 
\end{eqnarray}
Here, $g$ is a Gaussian random number. It is renewed at every time
step, $\tau_{h}$. The $\tau_{\xi}$ is the relaxation time of $\xi_{\alpha}$.
The parameters $\xi_{1,2,3}$ are introduced to model smooth variations
of $\xi_{\alpha}$. Similarly to the above-cited papers, we choose
the parameters of the Gaussian process as follows, $\overline{g}=0$,
$\sigma\left(g\right)=1$ and $\tau_{\xi}=2$ months. Parameters $\xi_{\alpha}$
and $\xi_{\beta}$ vary independently in the Northern and Southern
hemispheres.

\subsection{Data-Driven BMR Model\label{subsec:Data-driven}}

In the data-driven models, we compute the radial position of the unstable
point using the maximum of product $\left|\overline{B}\right|I_{\beta}\left(r,\theta\right)$,
the condition $I_{\beta}\left(r_{m},\theta_{m}\right)>0$ and $r_{m}>0.86R$.
For the low half part of the convection zone, we use function $\xi_{\beta}^{(-)}$.
The latitudinal and longitudinal coordinates $\theta_{m}$ and $\phi_{m}$
in function $\xi_{\beta}$ are taken from the active region data base.
Similarly to the theoretical model, the BMR's size is controlled by
the parameter $m_{\beta}$, which is taken in the form: 
\[
m_{\beta}=\frac{2}{\sqrt{\mathrm{S_{a}}/10^{6}}},
\]
where $\mathrm{S_{a}}$ is the maximum observed area (in millionth
of the hemisphere) of the bipolar active regions. We exclude all regions
with $\mathrm{S_{a}}<50$.

The solar observations show a wide range of variations in the BMR's
emergence time and growth rate. Also, there are periods of simultaneous
emergence of several BMR in one hemisphere. To avoid the overlaps,
we reformat the emergence initiation time as follows. Firstly, for
such cases, we shift the emergence of subsequent BMR by 2 time steps
after the end of the previous BMR emergence. Secondly, we define the
minimal emergence time $\delta\mathrm{t_{min}}=2$ days and assume
that these BMR have smaller $\tau_{0}$ or the higher growth rate.
Specifically, we define: 
\[
\mathrm{\tau_{a}}=\frac{1}{2}\frac{\mathrm{\delta t_{a}}}{\delta\mathrm{t_{min}}}\tau_{0},
\]
where $\mathrm{\delta t_{a}}$ is the total emergence time of the
BMR, $\mathrm{\tau_{a}}$ is the growth rate.

Putting all together, we obtain the instability function for the data-driven
modeling, $\xi_{\beta}^{\mathrm{(a)}}$ 
\begin{eqnarray}
\!\xi_{\beta}^{\mathrm{(a)}}\left(\boldsymbol{r},t\right)\! & = & \!\!C_{\beta}\negthinspace\tanh\mathrm{\left(\frac{t}{\tau_{a}}\right)}\exp\left(-\frac{2}{\sqrt{\mathrm{S_{a}}/10^{6}}}\left(\sin^{2}\!\left(\!\frac{\phi-\phi_{m}}{2}\!\right)\right.\right.\!\nonumber \\
 &  & \left.\left.+\!\sin^{2}\!\left(\!\frac{\theta-\theta_{\mathrm{m}}}{2}\!\right)\!\right)\right)\psi(r,r_{m}^{(+)}),t\!<\!\delta\mathrm{t_{a}}\label{xiba}\\
 & = & 0,t>\delta t_{a}\,.\nonumber 
\end{eqnarray}
The parameters $S_{a}$, $\theta_{m}$ and $\phi_{m}$ are taken from
the NOAA database of solar active regions (https://www.swpc.noaa.gov/).

\subsection{Boundary conditions}

The model divides the integration domain into two parts. The overshoot
region includes the upper part of the radiative zone. The bottom of
the integration domain is fixed at $r_{i}=0.67$R. The convection
zone extends from $r_{b}=0.728R$ to $r_{t}=0.99R$. The bottom boundary
rotates as a solid-body at the rate $\Omega_{0}=430$~nHz. At the
bottom boundary, the magnetic field induction vector is zero. At the
top boundary, we use the black-body radiation heat flux and the stress-free
condition for the hydrodynamic part of the problem.

For the induction equation (\ref{eq:mfe}), we use the top boundary
condition in the form that allows penetration of the toroidal magnetic
field to the surface: 
\begin{eqnarray}
\delta\frac{\eta_{T}}{r_{\mathrm{t}}}B\left(1+\frac{\left|B\right|}{B_{\mathrm{esq}}}\right)+\left(1-\delta\right)\mathcal{E}_{\theta} & = & 0,\label{eq:tor-vac}
\end{eqnarray}
where $r_{\mathrm{t}}=0.99R$. For the set of parameters $\delta=0.999$
and $B_{\mathrm{esq}}=5$G, the surface toroidal field magnitude is
around 1.5~G. {In compare to the standard case of the vacuum
boundary condition, the Eq( \ref{eq:tor-vac}) condition provides
a closer penetration of the toroidal dynamo wave toward the surface
and support the equatorward propagation of the toroidal magnetic field
due to the the Parker-Yoshimura rule (\citealt{Yoshimura1975}). This
can impact the BMR productivity at the near surface levels. Nothyworthy,
the solar observations indicate the surface toroidal field magnitude
is around 1G (\citealt{Vidotto2018}).} The poloidal magnetic field
is potential outside the dynamo domain. For the numerical solution,
we use a spectral expansion in terms of the spherical harmonics and
employ the \textsc{Fortran} version of the \textsc{shtns} library
of \citet{shtns}.

\section{Results}

\subsection{BMR formation and their parameters}

Table\ref{tab2} summarizes the key parameters of the model runs.
\begin{deluxetable*}{cccccc}
\tablenum{2}
\tablecaption{\label{tab2}The parameters of the model runs}
\tablewidth{0pt}
\tablehead{
\colhead{Model} & \colhead{BMR injection} & \colhead{$\alpha_{\beta}$}
&\colhead{$C_{\alpha}$} & \colhead{t, [yr]} & \colhead{Period,[yr]}
}
\decimalcolnumbers
\startdata
T0 & 0 & 0 & 0.045 & - & 10.4\\
T1 & $\xi_{\beta}^{(-)}$, $\xi_{\beta}^{(+)}$ & $\xi_{\alpha}$ & 0.045
& $\ge$0 & 10.6, 10.8, 10.5\\
T2 & $0$, $\xi_{\beta}^{(+)}$ & $\xi_{\alpha}$ & 0.04 & $\ge$0 & 11.2,
11.3,11.1\\
S0 & $\xi_{\beta}^{(-)}$, $\xi_{\beta}^{(a)}$ & $\xi_{\alpha}$ & 0.045
0.035, 0.044 & $\ge$0, $\ge$5, $\ge$11 & 11.2, 11.6\\
S1 & $\xi_{\beta}^{(-)}$, $\xi_{\beta}^{(a)}$ & $\xi_{\alpha}$
$\ge$0.95R & 0.045, 0.035,  0.044 & $\ge$0, $\ge$5, $\ge$11 & 11.4, 11.6\\
S2 & 0, $\xi_{\beta}^{(a)}$ & $\xi_{\alpha}$ & 0.045, 0.035, 0.04 & $\ge$0
$\ge$5, $\ge$11 & 11.4, 12\\
\enddata
\tablecomments{T0 is the axisymmetric
base model without BMR. T1 and T2 are models with the random initialization
of BMR. S0-S2 are data-driven models with the initialization corresponding
to Solar Cycles 23 and 24 in the upper half of the convection zone.
The second column shows the implementation of the BMR perturbations
in the lower half (the first function) and the top half (the second
function) of the convection zone (see Eqs~\ref{xib} and \ref{xiba});
the third column shows whether the models employ the BMR tilt (see,
Eqs~\ref{eq:ab},\ref{xia}); the column $C_{\alpha}$ show the parameters
of the global mean-field alpha effects (see the text); the next column
shows the time intervals for the corresponding $C_{\alpha}$ values
(after the start of Solar Cycle 23 for models S0-S2); the last column
shows the duration of the activity cycles (half dynamo periods of
the magnetic cycles)}
\end{deluxetable*}

Model T0 is a base axisymmetric dynamo model without BMR. It reproduces
basic observed properties of the solar cycles, such as the magnetic
butterfly diagram, polar field reversals, migrating zonal flows (torsional
oscillations), variations of the meridional circulation, and the extended
solar cycle phenomenon \citep{Pipin2019c,Pipin2020}. Models T1 and
T2 include the BMR initiation driven by the magnetic buoyancy instability
with initial perturbations with the radius according to the instability
criterion and randomly in longitude and latitude. The perturbations
are initiated in the whole convection zone in model T1, and only in
the upper half of the convection zone in model T2. Models S0-S2 are
data-driven models. Like in T1 and T2 models, the BMR sources are
distributed with the radius according to the magnetic buoyancy criterion,
but in the upper half of the convection zone the latitudinal and longitudinal
distributions correspond to the location of the solar active regions
observed during Solar Cycles 23 and 24 (BMR injection function $\xi_{\beta}^{(-)}$).
In the lower half of the convection zone, the BMR injection function
is random in longitude and latitude ($\xi_{\beta}^{(-)}$) models
S0 and S1, and it is not included in model S2. Also, in model S1,
we limit the BMR $\alpha$-effect to a near-surface layer. Then, the
BMR remain untilted deep in the convection zone.

The global $\alpha$-effect parameter, $C_{\alpha}$, is chosen to
match the duration and strength of the solar cycles. In particular,
to fit the parameters of the sunspot cycles, 23 and 24, in the data-driven
models S0 and S1 to observations, we use the variable mean-field $\alpha$-effect
because these cycles have different magnitudes and durations. The
data-driven models start from the epoch of the solar minimum at the
beginning of the sunspot cycle 23 in 1996. Using the numerical experiments,
we find that the prolonged decay of Cycle 23 can be modeled if parameter
$C_{\alpha}$ is decreased by 20\% relative to its initial value $C_{\alpha}=0.036$
in 2001, after five years from the start of Cycle 23. Then, at the
end of the cycle in 2011, it is increased back by 15\% to $C_{\alpha}=0.04$.
The value $C_{\alpha}=0.04$ is close to the dynamo threshold. The
same variations of $C_{\alpha}$ are used in model T1. The reference
axisymmetric model T0 has the constant $C_{\alpha}=0.045$.

\begin{figure}
\centering \includegraphics[width=0.7\columnwidth]{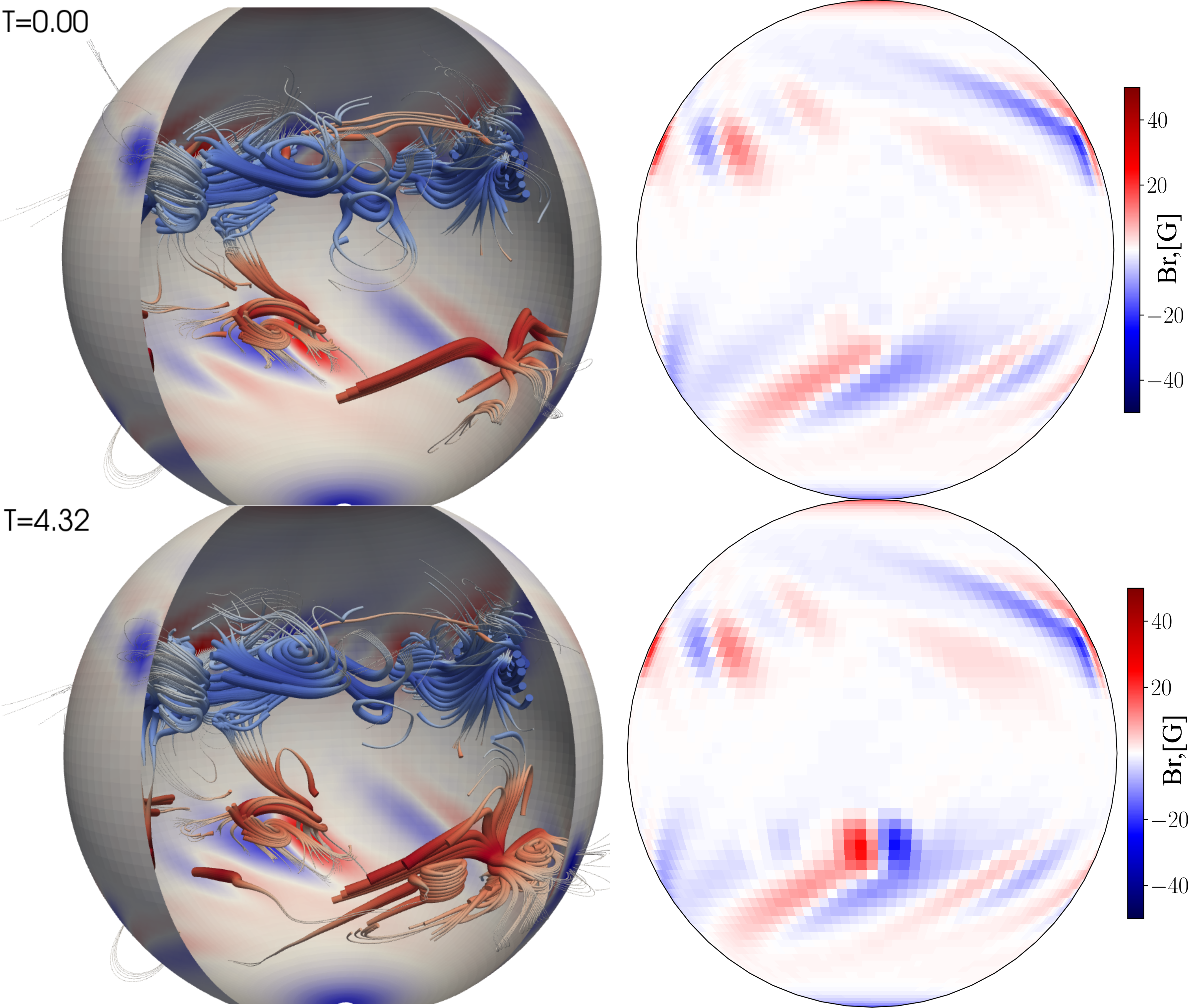} \caption{\label{fig:form} Formation of two magnetic regions in the South hemisphere
during the growth phase of Cycle 23 in model S0. Left column shows
the field lines of the non-axisymmetric magnetic field, and time is
shown in days. Right column shows the surface radial magnetic field.
{The full animation of the magnetic field evolution is available
online. The animation illustrates the formation of BMR in the South
hemisphere. The BMR start as magnetic bubbles which eventually appear
at the surface. Also, we see effect of the large-scale flow and the
eddy magnetic diffusivity on evolution of the BMR's remnants.}}
\end{figure}

Figure \ref{fig:form} illustrates the formation of BMR simulated
in model S0. The BMR start as magnetic bubbles which eventually appear
at the surface. In addition, we see the corresponding poleward magnetic
flux transport events that are formed from remnants of the BMR's evolution
due to the effects of differential rotation, meridional circulation,
and magnetic eddy diffusivity.

\begin{figure}
\centering \includegraphics[width=0.7\columnwidth]{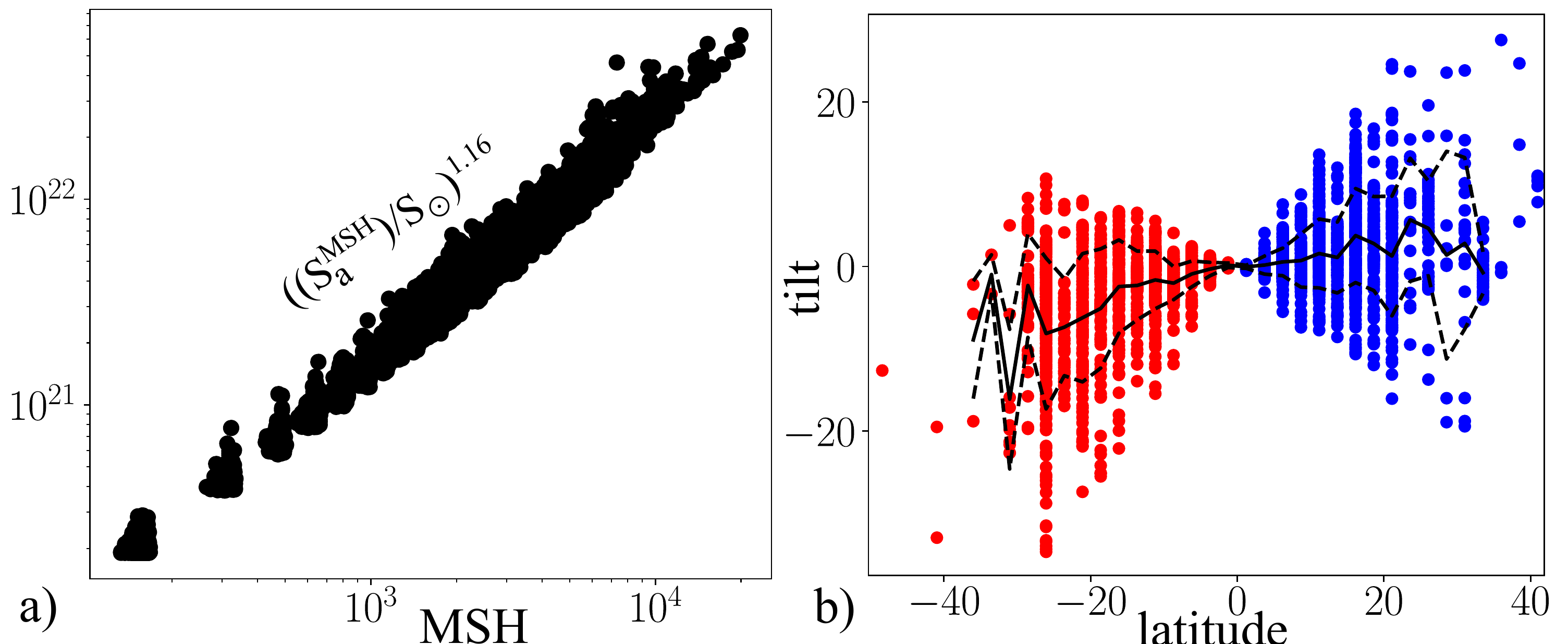} \caption{\label{fig:fla} Model S0: a) Distribution of the BMR area and the
magnetic flux, b) the BMR's tilt. Blue color shows BMR from the northern
hemisphere, and red color shows the southern hemisphere, the solid
line shows the mean tilt, and the dashed lines show the tilt variance.}
\end{figure}

\begin{figure}
\centering \includegraphics[width=0.7\columnwidth]{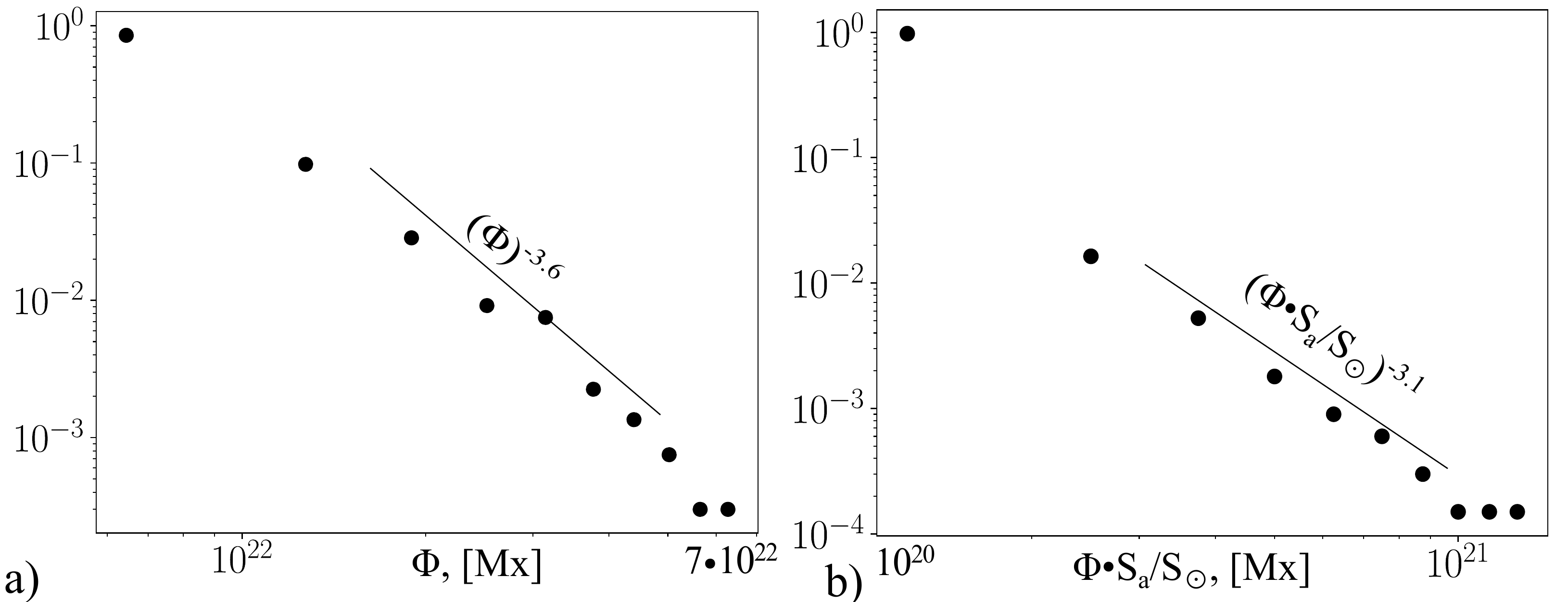} \caption{\label{fig:freq} Model S0. a) Occurrence probability for the BMR
flux magnitude, $\Phi$; b) the same for the BMR flux and area magnitude,
$\Phi\cdot\mathrm{S_{a}}$}
\end{figure}

Next, we look at distributions of the BMR's size, tilt angles and
magnetic fluxes. We consider snapshots of the synoptic maps of the
non-axisymmetric radial magnetic field and calculate the continuous
area of the magnetic regions using the threshold of $10^{20}$ Mx
in the pixel. We find the linear relation between the BMR's area and
flux, see Fig.~\ref{fig:fla}a). This is in agreement with observational
results \citep{Nagovitsyn2021}. The distribution of the tilt angle
shown in Fig.~\ref{fig:fla}b is also in close agreement with observations
\citep{Nagovitsyn2021M}, including the nonlinear behavior at latitudes
$>25^{\circ}$.

Figure~\ref{fig:freq} shows results for the probability distributions
of the BMR's flux magnitude, $\Phi$, and the power of the magnetic
flux occupying the area $\mathrm{S_{a}}$, $\Phi\cdot\mathrm{S_{a}}$.
These parameters show the power-law probability distributions, similarly
to the observational analyses of \citet{Parnell2009}, \citet{Munoz2015}
and \citet{Nagovitsyn2021M}. However, the power-law indexes of in
models are higher than in the observations, e.g., \citet{Parnell2009}
found $P\propto\left(\Phi\cdot\mathrm{S_{a}}\right)^{-1.8}$. Our
model reproduces the power law but with a steeper index $\approx-3.1$.
\begin{figure}
\centering \includegraphics[width=0.7\textwidth]{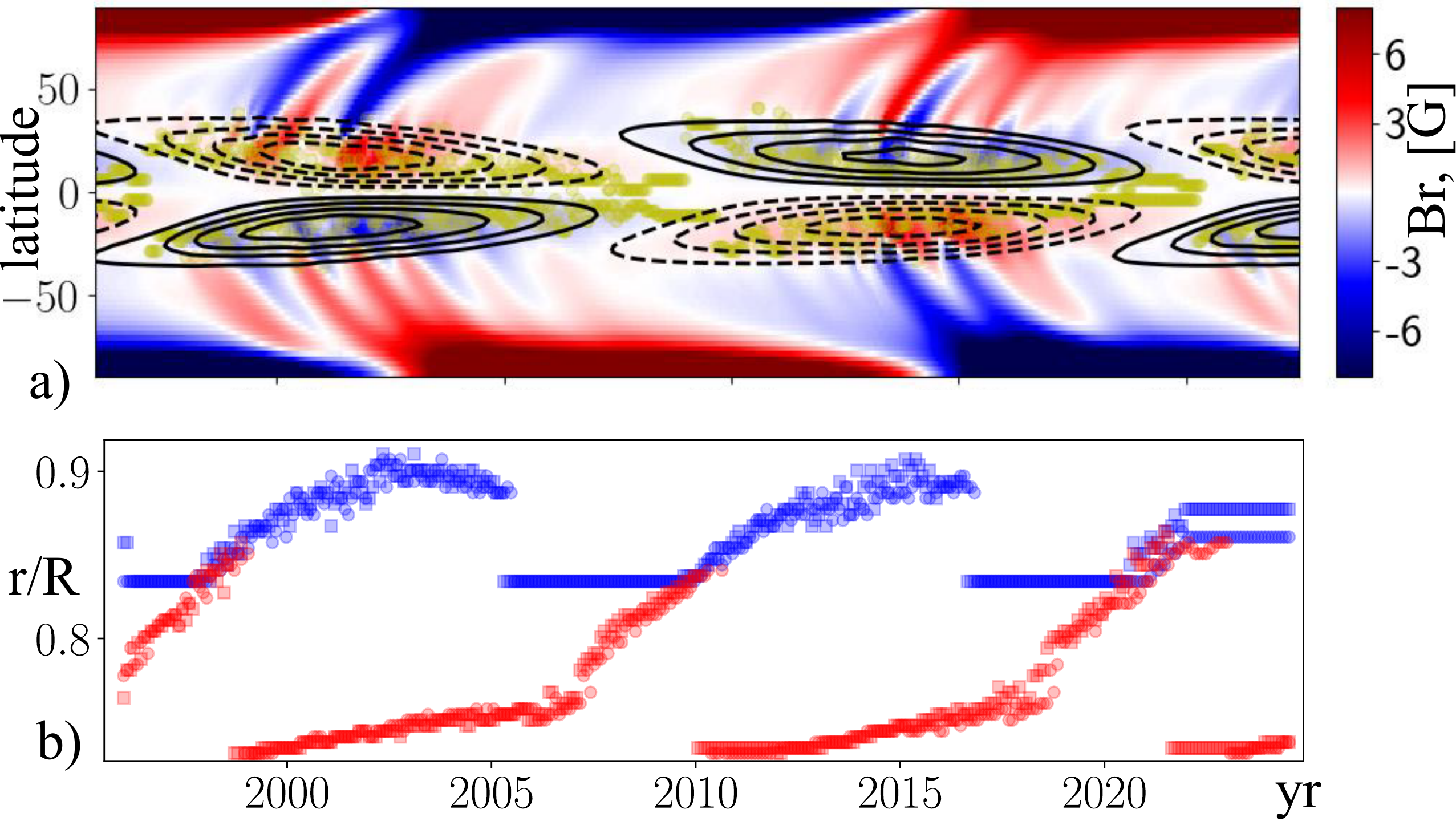} \caption{\label{fig:s0} Model S0. a) Time-latitude diagram of the near surface
toroidal magnetic field (contours in the range of $\pm1$kG), surface
radial magnetic field (color image), and positions of the BMR (squares)
for model S0; b) the starting radial positions of the injected BMRs;
blue squares mark the BMR injections from the upper part of the convection
zone and the red squares mark the injections from the lower part.}
\end{figure}

\begin{figure}
\centering\includegraphics[width=0.5\textwidth]{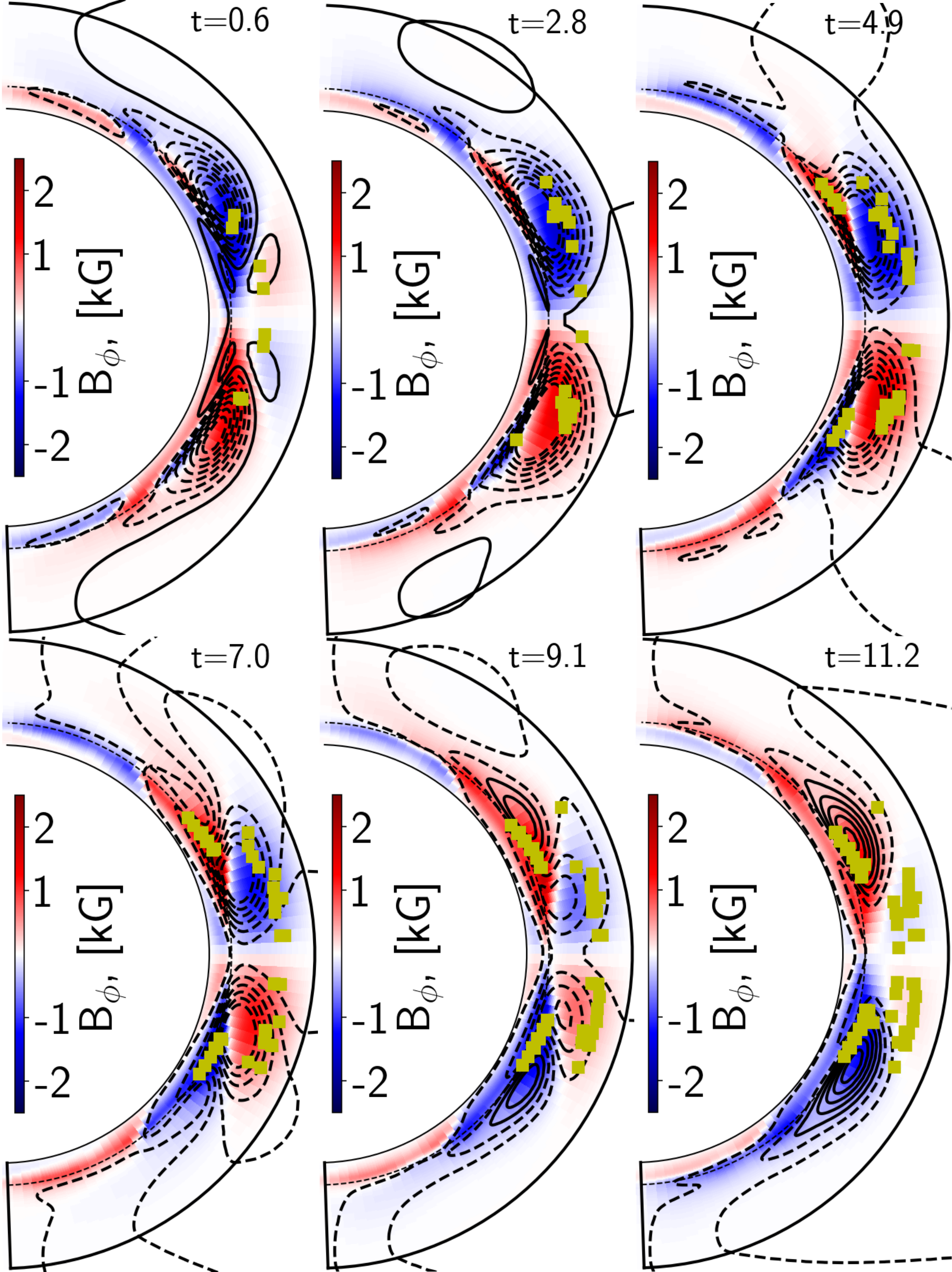} \caption{\label{fig:s0sn} Snapshots of the axisymmetric magnetic field evolution
during Cycle 23 in model S0. The BMR formation locations are marked
by squares. To show the weak toroidal field near the equator, we use
over-saturated color range. Vector-potential streamlines show the
poloidal magnetic field, dashed lines reflect the anticlockwise direction
of the field lines.}
\end{figure}

\begin{figure}
\centering \includegraphics[width=0.7\textwidth]{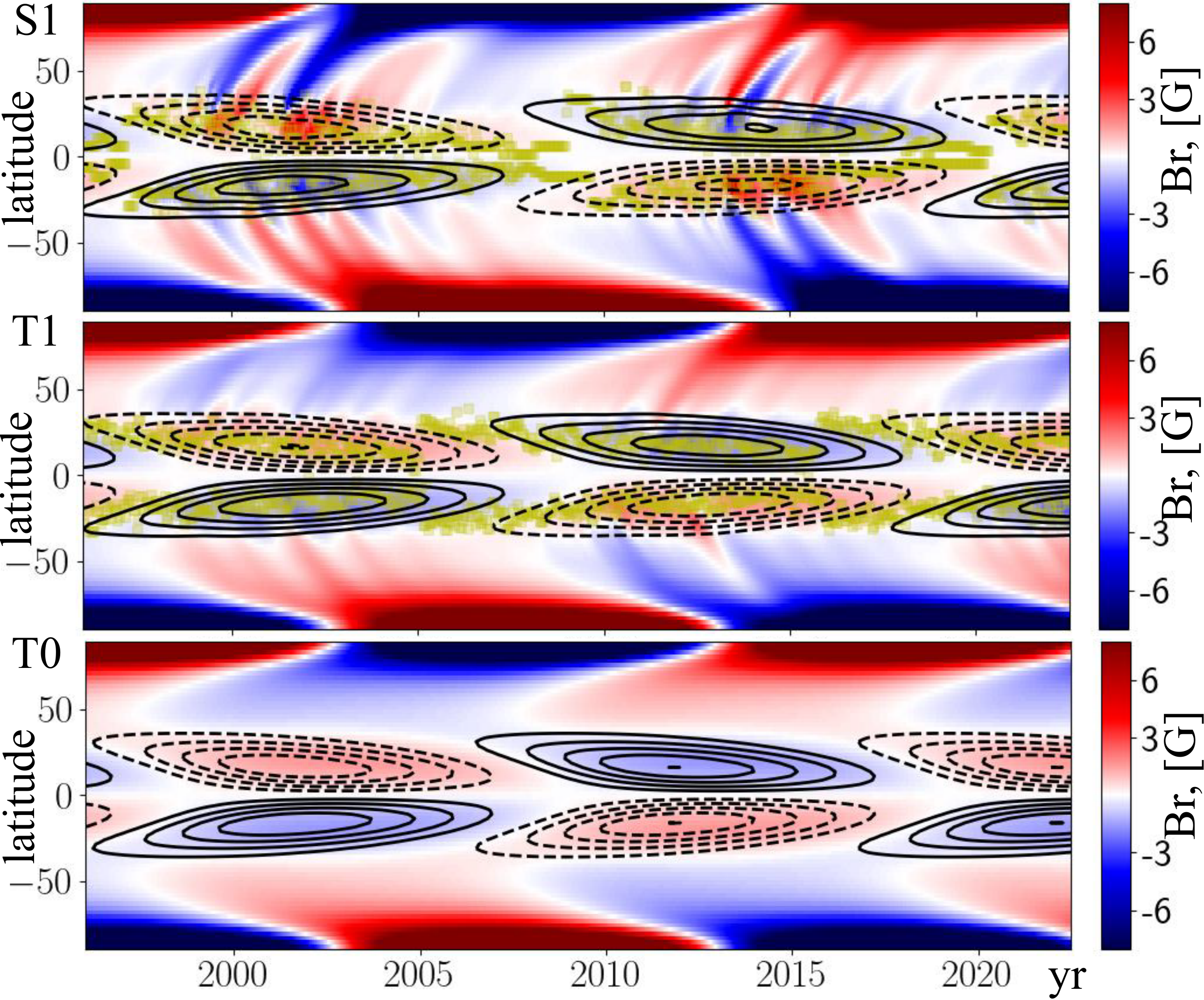} \caption{\label{fig:s1tl} a) Time-latitude diagram of the near surface toroidal
magnetic field (contours in range of $\pm1$kG), surface radial magnetic
field (color image), and positions of the BMR (squares) for model
S1; b) and c) the same as a) for models T1 and T0.}
\end{figure}

\subsection{The modeled magnetic field evolution}

Figure \ref{fig:s0} shows the time-latitude diagram of the surface
radial magnetic field and the toroidal magnetic field in the subsurface
shear layer for model S0. Also, we show the evolution of the BMR injection
locations in radius. The butterfly diagrams are similar to the results
of P22. The radial locations of the BMR injections mark the propagation
of the dynamo wave from the bottom of the convection zone toward the
surface \citep{Kosovichev2019,Pipin2019c}. Figure \ref{fig:s0sn}
illustrates this propagation in a series of magnetic field meridional
snapshots.

Noteworthy, the BMR injections from the bottom of the convection zone
do not produce the surface BMR. This may be related to the restricted
radial sizes of the BMR initiation sources (see Eq.~\ref{eq:step}).
Nevertheless, as discussed later, such magnetic flux injections affect
the surface non-axisymmetric magnetic field. With the described tuning
of the $C_{\alpha}$ parameter, the BMR activity in model S0 satisfactorily
fits the time-latitude variations of the near-surface toroidal magnetic
field. Yet, near the equator, the BMR activity goes outside the modeled
toroidal field evolution during the epoch of Cycle 23 minimum. The
situation is clarified by Fig.~\ref{fig:s0sn}. During the activity
minima, the magnetic buoyancy mechanism initiates BMR not only from
weak remnants of the toroidal magnetic field of the old cycle in the
subsurface layers very near the equator but also from the deeper layers
on the edge of the new dynamo wave of the next cycle.

Figure~\ref{fig:s1tl} shows the time-latitude diagrams for models
S1, T0 and T1. Similarly to th e results of P22, we find that the
model produces the smooth evolution of the surface radial magnetic
field if we neglect the BMR's tilt. The comparison of models S0 and
T1 shows that details of the magnetic buoyancy mechanism and latitudinal
locations of the BMR activity are important for the magnetic cycle
parameters \citep{Mackay2012,Miesch2014Ap}. For comparison, we show
the results for the pure axisymmetric model, T0, as well. For the
given $C_{\alpha}$, model T0 has the activity cycle (half of the
full dynamo period) of about 10.5 years, which is shorter than the
periods of models T1 and S0, which include the BMR activity. %$\overline{\left|\left\langle \mathbf{B}_{r}\right\rangle \right|}$,

\begin{figure}
\centering \includegraphics[width=0.6\columnwidth]{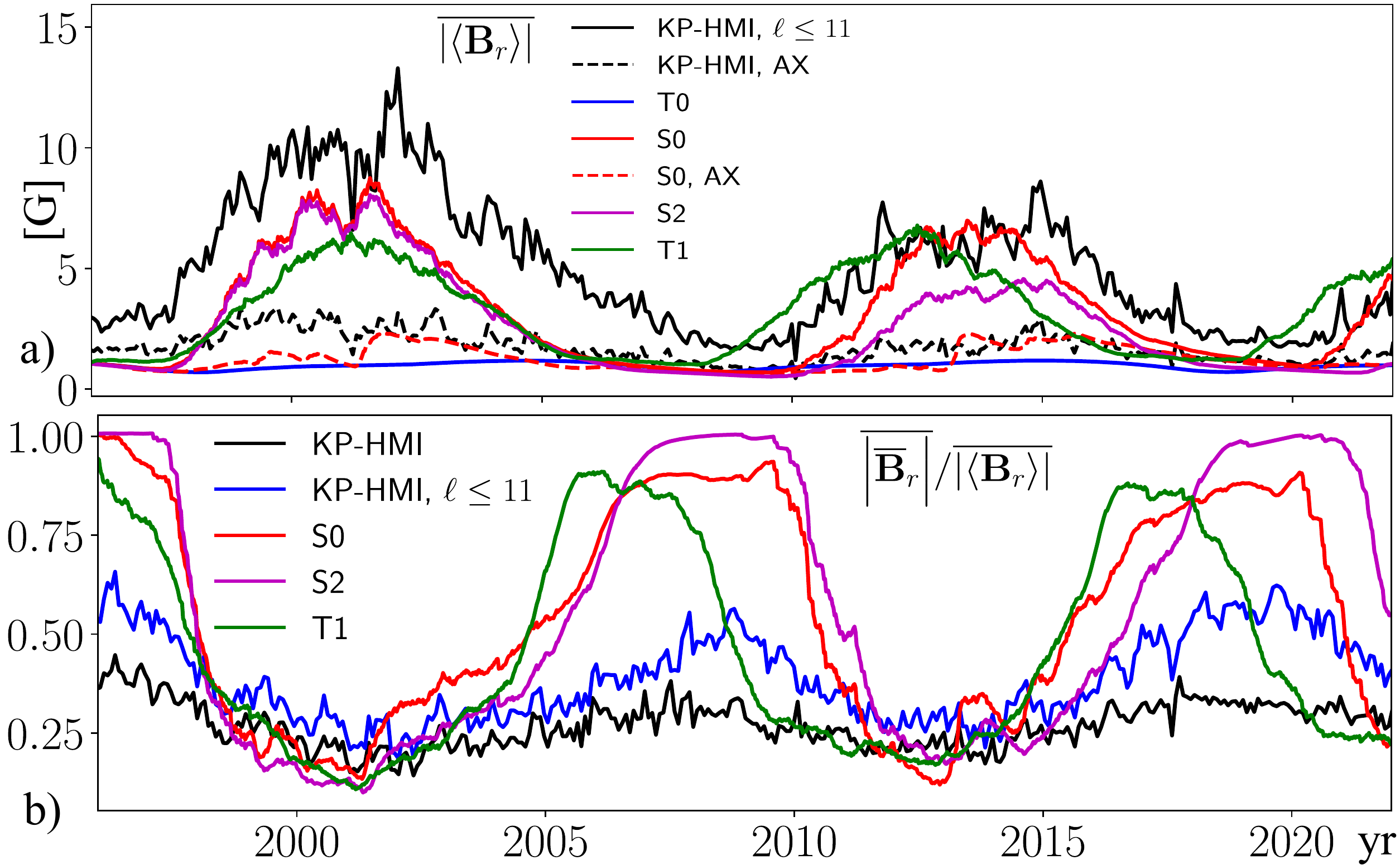} \caption{\label{fig:s1fl}a) The mean absolute magnitude of the surface radial
magnetic field; b) the ratio of the mean absolute magnitude of the
axisymmetric surface field component to the mean magnitude of the
surface radial magnetic field, $\overline{\left|\overline{\mathbf{B}}_{r}\right|}/\overline{\left|\left\langle \mathbf{B}_{r}\right\rangle \right|}$.}
\end{figure}

Figure~\ref{fig:s1fl} shows the mean absolute magnitude of the surface
radial magnetic field, $\overline{\left|\mathbf{B}_{r}\right|}$,
and the ratio of the mean magnitude of the axisymmetric surface field,
${\left|\overline{\mathbf{B}}_{r}\right|}$ to $\overline{\left|\mathbf{B}_{r}\right|}$.
This ratio characterizes the level of the non-axisymmetry of the surface
activity in our models. To compare with observations, we use the synoptic
maps of the radial magnetic field from the KPO, SOLIS and SDO/HMI
data archives \citep{Harvey1980,Bertello2014,Scherrer2012}. Using
the synoptic maps, we calculate the surface mean of the unsigned radial
magnetic field and the same for the axisymmetric radial magnetic field.
We find that, in the observations, the value of $\overline{\left|\mathbf{B}_{r}\right|}$
reaches about 20-25~G at the solar maxima, and it was around 15~G
in Solar Cycles 23 and 24. Bearing in mind the large-scale character
of the magnetic activity in our model, we also compare our results
with $\overline{\left|\mathbf{B}_{r}\right|}$ calculated f or non-axisymmetric
spherical harmonics of the angular order $m<11$. We find a satisfactory
agreement of models S0 and S2 with the solar observations. Model S0
shows that Solar Cycle 25, started in 2019, can be the same or a little
higher than Cycle 24. The same is likely true in model S2.

However, the basal level of $\overline{\left|\mathbf{B}_{r}\right|}$
during the solar minima is about by a factor of 2 smaller than in
the solar observations. This is reflected in the behavior of the non-axisymmetry
parameter ${\left|\overline{\mathbf{B}}_{r}\right|}/\overline{\left|\mathbf{B}_{r}\right|}$,
as well. It seems that our model misses some parts of the dynamo process
that are essential for the large-scale non-axisymmetric magnetic field
of the Sun. Interestingly, the contribution of the BMR activity to
the axisymmetric magnetic field increases to the observational level
just after the magnetic cycle maximum (see the dashed red curve in
Fig.~\ref{fig:s1fl}a). Model S2 shows the same behavior. 

\begin{figure}
\centering \includegraphics[width=0.6\columnwidth]{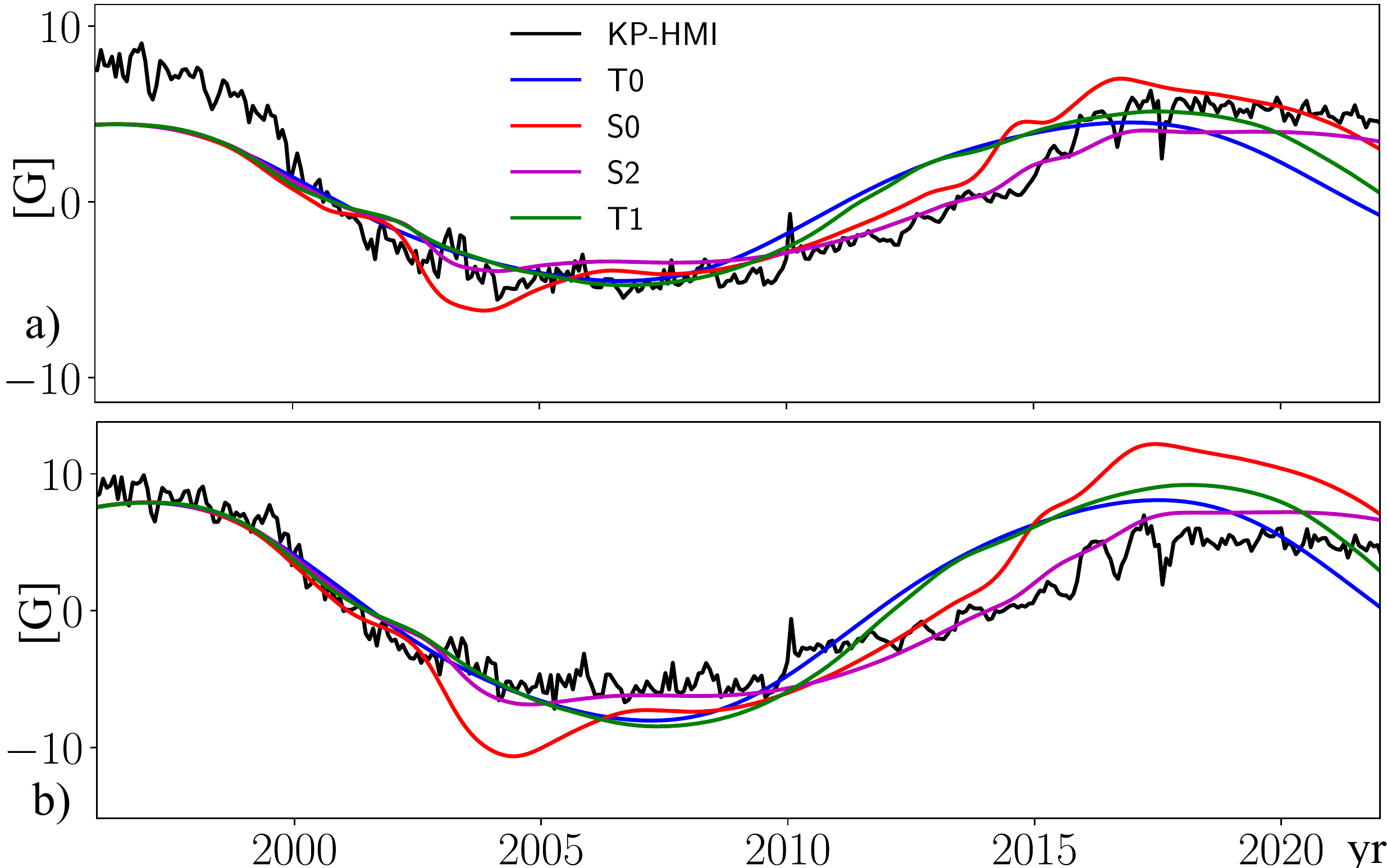} \caption{\label{fig:pol} a) The mean axisymmetric radial magnetic field above
$60^{\circ}$ latitude at the surface; b) the same above $70^{\circ}$
latitude.}
\end{figure}

Figure~\ref{fig:pol} shows the evolution of the polar magnetic field
calculated by averaging the North-South anti-symmetric radial field
components for latitudes higher than $60$ and $70$ degrees. We see
that the polar field in the solar observations shows only a little
change between these two measurements during the epoch of solar minimum
around 1996. At the same time, the dynamo model shows significant
changes in the latitude range in the polar field definition. Model
T0 without BMR shows the same magnitude of the polar field for the
cycle minima in 1996, 2007, and 2018. Model T1, with random initialization
of BMR, shows a slight increase of the polar magnetic field during
the minima due to the BMR contribution to the polar field magnitude.
The same effect is found for model S0; although the magnitude of the
mean-field $\alpha$-effect in this model was decreased in Cycles
23 and 24 to match the observed properties of Cycle 24. The mean latitude
of the BMR injections in the data-driven model S0 (NOAA data driven
model) is lower than in model T1 with the random BMR initialization.
Model S2 shows the best fit for the solar data reproducing plateaus
during the minima of Solar Cycles 23 and 24.

In our simulations, we investigated various possibilities to reproduce
the basic parameters of Solar Cycles 23 and 24. Besides the long-term
variations of the mean-field parameter $C_{\alpha}$, we considered
$\alpha_{\beta}$ variations. In such cases, to bring the model in
the best agreement with observations, we needed to assume very small
values $C_{\beta}$ (corresponding to the mean BMR tilt) during the
declining phase of Cycle 23 and the growth phase of Cycle 24. However,
the observational results of \citet{Tlatov2013} do not show strong
variations of the mean BMR tilt in different solar cycles. Yet, we
can not exclude that Cycles 23 and 24 were affected by the emergence
of the so-called ``rogue'' active regions \citep{Nagy2017,Kumar2021}.
This point should be studied separately.

\subsection{Torsional oscillations and meridional circulation variations}

\begin{figure}
\includegraphics[width=0.99\columnwidth]{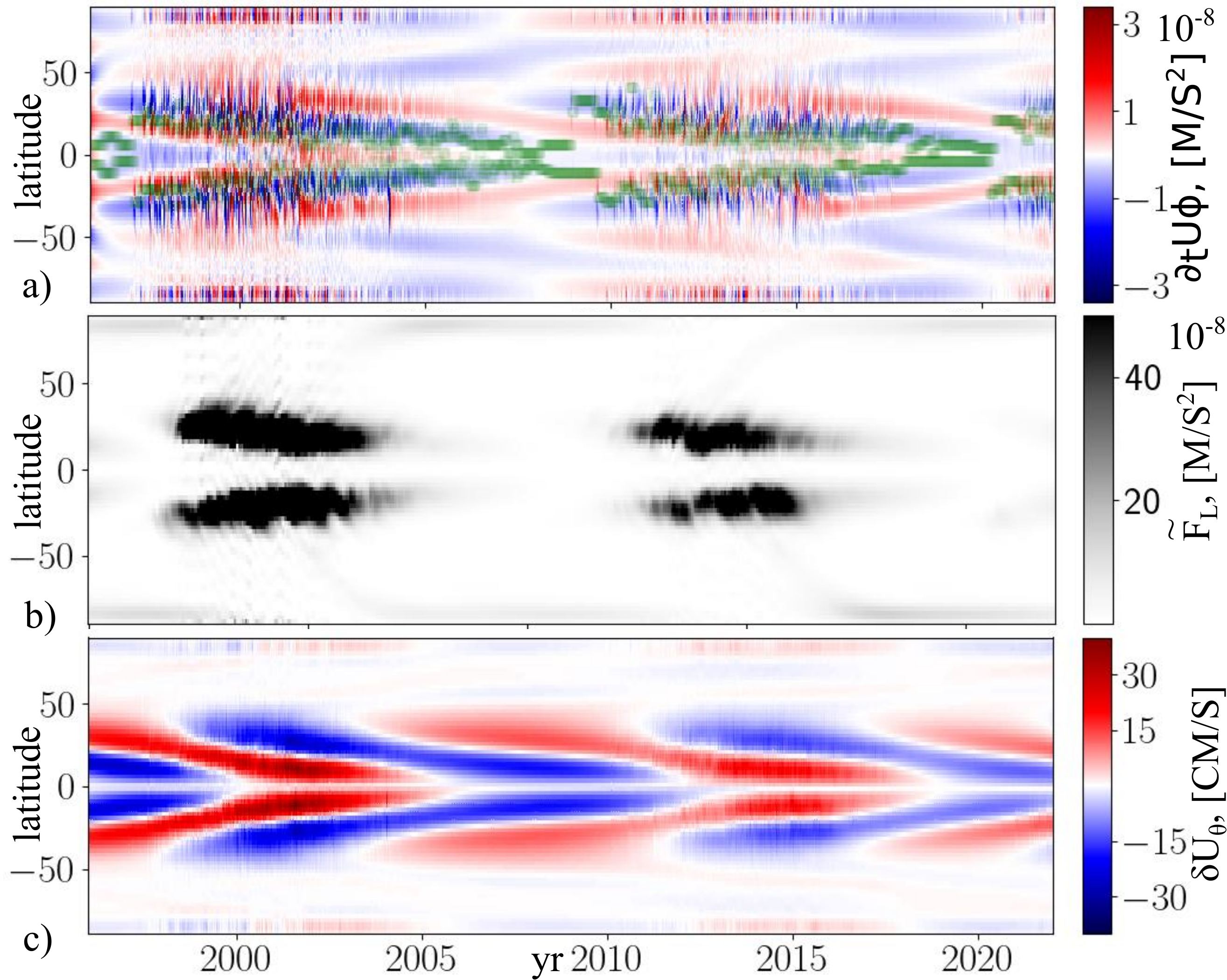}

\caption{\label{fig:to}a) The background blue-red color show the surface variation
of the zonal acceleration, the BMR activity latitude is shown by green
squares; b) The time-latitude variations of the azimuthal force induced
by the BMR activity; c) the time-latitude variations of the surface
meridional circulation, the positive variation is toward the poles. }
\end{figure}

{We find the BMR activity makes a substantial contribution in variation
of the zonal flow and meridional circulation at the surface. The Fig\ref{fig:to}
show the surface time-latitude diagrams of the zonal acceleration,
the Lorentz force from the BMRs and variations of the meridional circulation for the run S2. }

{The Lorentz force induced by the BMR's activity, $\tilde{F}_{L}$,
is determined by the azimuthal average of the magnetic stress as follows:
\begin{eqnarray}
\tilde{F}_{L} & = & \frac{1}{\overline{\rho}r\sin\theta}\boldsymbol{\nabla\cdot}\left(r\sin\theta\frac{\tilde{\mathbf{B}}\tilde{B}_{\phi}}{4\pi}\right).\label{flp}
\end{eqnarray}
Similar to results of \citet{Pipin2019c}, the model show the extended cycle of the torsional oscillation: the acceleration wave which started at about 50$^{\circ}$ latitude around 1997 finished at equator after 20 years. The time-latitude pattern of the torsional oscillation results from a complicate force balance. The angular momentum balance includes, the dynamo induced variations of turbulent stresses, inertia forces,
and variations of the meridional circulation, see a detailed study
in the above cited paper. Noteworthy the amplitude of the forces which contribute the balance is more than order of magnitude higher than the magnitude of the zonal acceleration. Here, we see that the BMR activity produce the azimuthal acceleration of the near equatorial regions of the Sun.}

{The model shows the North-South asymmetry of the meridional circulation variations in the cycle 24. This asymmetry is accompanied by the trans-equatorial meridional flow of small magnitude during the growing phase of the cycle 24. Similar results was found recently by \citet{Getling2021}
from helioseismology. In Fig\ref{fig:to}c we saturate the variations of high magnitude to show the weak variations of the meridional circulation in the polar regions. We see the decrease the meridional flow in the range of latitudes from 30$^{\circ}$ to 50$^{\circ}$ and from 60$^{\circ}$
to 70$^{\circ}$ during epochs of the magnetic activity maximum. The high latitude variations of the meridional circulation can affect the surface magnetic flux transport and evolution of the polar magnetic field.}

\section{Discussion and Conclusions}

We model the physical parameters of Solar Cycles 23 and 24 using the
nonlinear 3D mean-field dynamical dynamo model and the observational
active region data. Our algorithm for the emergence of bipolar magnetic
regions is based on the magnetic buoyancy effect acting on the unstable
part of the large-scale magnetic field. The radial positions of the
unstable regions are calculated using Parker's magnetic instability
condition. For the dynamo process distributed in the convection zone,
this condition leads to the instability of the toroidal magnetic field
at the front edge of the dynamo waves near the bottom of the convection
zone and in its upper part, as illustrated in Fig.~\ref{fig:s0sn}.
For modeling the solar BMR injected from the upper part of the convection
zone, we use the NOAA data base for coordinates and areas of active
regions to specify the locations and sizes of the initial perturbations.
For the unstable regions in the lower part of the convection zone,
we use perturbations randomly distributed in time and longitude. Our
results show that, in most cases, magnetic flux injections from the
lower part of the convection zone do not result in the BMR formation
on the surface. However, these injections affect the magnitude of
the background non-axisymmetric magnetic field on the surface. Our
dynamo models show the basal level of the non-axisymmetric magnetic
field during the solar minima is about a factor of two smaller than
in the solar observations. The possible reason is that our BMR algorithm
is not sufficient for deep initialization sources. This affects the
onset and decay of the large-scale non-axisymmetric magnetic activity.

The BMR distributions generated by our dynamo models show that the
magnetic flux is directly proportional to BMR's areas. This result
is equally applied to the data-driven models S0, S1, and S2, and models
T1 and T2 with the random distribution of the BMR sizes and longitudinal
initialization points. The same proportionality was found in observations
by \citet{Nagovitsyn2021} for the sunspot groups areas. The distribution
of the BMR vs. the magnetic flux and area shows the inverse power-law
of index -3.1. This qualitatively agrees with the results of \citet{Parnell2009},
\citet{Munoz2015}, and \citet{Nagovitsyn2021M}. However, they found
a less steep power law for the large-scale part of the magnetic field
distributions.

In our models, the BMR's tilt is given theoretically, and it is directly
related to the near-surface $\alpha$-effect \citealp{Stix1974,Pipin2022}.
The latitudinal profile of the tilt follows the $\cos\theta$ dependence,
where $\theta$ is co-latitude. We choose the BMR's tilt to be randomly
fluctuating about the mean. The resulting tilt distribution agrees
with the solar observations \citep[e.g.][]{Nagovitsyn2021M}. Also,
these authors found a tendency for the nonlinear behavior of tilt
at latitudes $>25^{\circ}$. Our model shows similar behavior. It
is caused by the $\alpha$-effect modulation due to the large-scale
toroidal magnetic field.

{In general, the dynamo solution of our model follows the Parker-Yoshimura
law (\citealt{Pipin21c}). In the weakly nonlinear regime the dynamo
period of the Parker-Yoshimura dynamo waves is linearly connected
with the $\alpha$ effect magnitude (\citealt{Noyes1984}). Naturally,
the distributed dynamo models demonstrate the Waldmeier rules, as
well (\citealt{Pipin2011}). We exploited these properties to model
the prolonged cycle 23 using the temporal decrease of the turbulent
$\alpha$-effect. Generally, we can not exclude the other reasons
for unusual behavior of cycle 23. In particular, \citet{Dikpati2010G},
using the flux-transport model, argued that the long cycle 23 can
be due to increase of the meridional circulation cell. They suggested
that in the standard situation the solar dynamo is controlled by two
meridional circulation cell per hemisphere , and the ``polar'' cell
has the inverted circulation in compare to the main ``equatorial''
cell.  For the strongly nonlinear
case the model can reproduce the multiple cell circulation pattern. Results of \citet{Pipin21c}
showed that this can happen on the fast rotating solar analogs. The
dynamo period in this case is about 2 year. This is substantially
less than the solar cycle period. It is an open question whether the
model can reproduce the multiple cell in latitude for the nonmagnetic
case.  \cite{Pipin2019c} found that for the solar-type dynamo the model shows a decrease of the surface poleward flow at high latitude during the magnetic activity maxima. Similar results are suggested by the helioseismology analysis of \cite{Getling2021}. Here, we demonstrated this effect in the modelled cycles 23 and 24. Simultaneously, we find that the hemispheric asymmetry of the sunspot activity during the growing phase of cycle 24 results in the North-South asymmetry of the meridional circulation variations. This asymmetry is accompanied by the trans-equatorial meridional flow of small magnitude during that epoch.
The variations of the meridional circulation are closely related with the torsional oscillations. These variations show the North-South asymmetry during the modelled cycle 24, as well. The model shows that the BMR activity induced the azimuthal acceleration force toward equator. It has the same order of magnitude as the other sources of the torsional oscillations due to the large-scale dynamo induced variations of turbulent stresses, inertia forces, and variations of the meridional circulation, (see, \citealp{Pipin2019c}).}

Similar to \citet{Pipin2021b} and \citet{Pipin2022}, we conclude
that our initial modeling of Solar Cycles 22 and 23, which includes
a combination of the global mean-field dynamo and emerging bipolar
magnetic regions (BMR), shows that the BMR activity plays a significant
role by affecting the strength and duration of the solar cycles. It
was missing in the previous Parker-type dynamo models of the solar
cycle. However, the data-driven models show that the BMR effect alone
cannot explain the weak Cycle 24. The decrease in the cycle amplitude
and the prolonged preceding minimum were probably caused by a decrease
of the turbulent helicity in the bulk of the convection zone during
the decaying phase of Cycle 23.

\textbf{Acknowledgments}

VP and VT thank the financial support of the Ministry of Science and
Higher Education of the Russian Federation (Subsidy No.075-GZ/C3569/278).
AK thanks the partial support of NASA grants: NNX14AB70G, 80NSSC20K0602,
80NSSC20K1320, and 80NSSC22M0162.

Data Availability Statements. The data underlying this article are
available by request.

\bibliographystyle{aasjournal}
\bibliography{dyn,dyn_bmr_2022}

\end{document}